\documentclass[aps,twocolumn,amsmath,amssymb,floatfix,showpacs]{revtex4-1}

\usepackage{graphicx}
\usepackage{dcolumn} 
\usepackage{bm}      
\usepackage{color}

\begin{document}

\title{Long-term memory in experiments and numerical simulations of hydrodynamic and
       magnetohydrodynamic turbulence}
\author{P. Mininni$^{1,2}$, P. Dmitruk$^1$, P. Odier$^3$, J.-F. Pinton$^3$, N. Plihon$^3$, G. Verhille$^4$, R. Volk$^3$, M. Bourgoin$^5$,}
\affiliation{$^1$Departamento de F\'\i sica, Facultad de Ciencias Exactas y
         Naturales, Universidad de Buenos Aires and IFIBA, CONICET, Ciudad
         Universitaria, 1428 Buenos Aires, Argentina. \\
             $^2$Computational and Information Systems Laboratory, NCAR,
         P.O. Box 3000, Boulder, Colorado 80307-3000, USA. \\
             $^3$Laboratoire de Physique - UMR 5672\\
             Ecole Normale SupŽrieure de Lyon / CNRS\\
             46 All\'ee d'Italie, 69007 Lyon, France\\
             $^4$ Aix-Marseille Universit\'e, IRPHE - UMR 7342, CNRS, Marseille, France\\
             $^5$ Laboratoire des Ecoulements G\'eophysiques et Industriels - UMR 5519\\
             Universit\'e Joseph Fourier / Grenoble-INP / CNRS\\
             BP53, 38041 Grenoble cedex 9.
}

\date{\today}

\begin{abstract}
We analyze time series stemming from experiments and direct numerical
simulations of hydrodynamic and magnetohydrodynamic
turbulence. Simulations are done in periodic boxes, but with a
volumetric forcing chosen to mimic the geometry of the flow in the
experiments, the von K\'arm\'an swirling flow between two
counter-rotating impellers. Parameters in the simulations are chosen
to (within computational limitations) allow comparisons between the
experiments and the numerical results. Conducting fluids are
considered in all cases. Two different configurations are considered:
a case with a weak externally imposed magnetic field, and a case with
self-sustained magnetic fields. Evidence of long-term memory and 
$1/f$ noise is observed in experiments and simulations, in the case 
with weak magnetic field associated with the hydrodynamic behavior of
the shear layer in the von K\'arm\'an flow, and in the dynamo case 
associated with slow magnetohydrodynamic behavior of the large 
scale magnetic field.
\end{abstract}
\maketitle

\section{Introduction}
Since the development of hot-wire measurements in hydrodynamic turbulence in
the laboratory, several reports of long-term memory in velocity time series (i.e.,
non-negligible correlations in time scales much larger than the time scale
associated with the energy containing eddies in the flow) can be found in the
literature (see, e.g., \cite{bib:delaTorre2007,bib:crespoDelArco2009_GAFD} for 
recent reports). The origin of such long-term correlations is unclear, as current 
knowledge of turbulence indicates that three-dimensional flows should transfer energy 
from the energy containing scale towards smaller scales (where the characteristic 
time scales are shorter). Therefore, no correlations at time scales longer than the
energy containing scale are to be expected.

Evidence of long-term memory (or ``long-range dependence'') is often observed
in the frequency spectrum of time series as a range of frequencies corresponding to 
$1/f$ noise. By $1/f$ noise (also often called ``flicker'' noise), it is usually 
meant that the power spectrum of the signal is of the form $E(f) \sim f^{-\alpha}$, 
where $f$ is the frequency, and with $\alpha$ loosely between $0.5$ and $1.5$. The 
case with $\alpha=1$, which strictly speaking corresponds to $1/f$ noise, is the 
case with equal energy per octave independent of the frequency. The range of 
frequencies in which this phenomenon often develops corresponds not only to 
frequencies smaller than the frequency associated with the energy containing 
eddies, but often also to frequencies that (if Taylor hypothesis are used) are 
associated with length scales much larger than the physical extension of the fluid. 
Therefore, such correlations cannot be associated with convective motions in the 
turbulent flow, and must be associated with long-term modulations in the system.

Observations of $1/f$ fluctuations are not exclusive of hydrodynamic or
magnetohydrodynamic (MHD) turbulence, and are widely found in natural and
nonlinear systems \cite{bib:Montroll1982}. As a result, they are often considered
as a signature of scale invariant features of an underlying dynamical process.
Time signals displaying $1/f$ noise in their spectral density have been reported
in electronics, tree growth, and human activities such as music and the stock 
market \cite{bib:Machlup1981}.
In conducting fluids and plasmas, $1/f$ fluctuations have also been reported
(see e.g., \cite{bib:ponty2004_PRL,bib:dmitruk2007_PRE}). In the interplanetary magnetic field, its
presence has been known for some time through analysis of time signals measured
{\it in situ} near Earth's orbit
\cite{bib:goldstein1995_ARA,bib:mattheaus1986_PRL,bib:matthaeus2007_AJ}.
Long-term behavior has also been found in time series of the geomagnetic field
\cite{bib:Carbone2006,bib:Ziegler2011,bib:Constable2007}, 
and in dynamo laboratory experiments \cite{bib:monchaux2009_PoF}. 

This ubiquity has motivated several works that
attempted to identify general sources of $1/f$ noise and long-term memory.
Recent studies focused in the particular cases of hydrodynamic and MHD flows
\cite{bib:dmitruk2007_PRE,bib:Dmitruk2011} obtained $1/f$ in MHD with and without 
a background magnetic field, but did not observe it in isotropic hydrodynamic 
flows. 
The modes giving the dominant contribution to the long-term correlations were 
identified to be modes with the largest available wavelength in the domain. 
Random isotropic forcing, with a short-time memory to give a single (unit) 
correlation time, was used to stir the flows in 
\cite{bib:dmitruk2007_PRE}, while random initial conditions in ideal systems 
were considered in \cite{bib:Dmitruk2011}. It was concluded that long-term memory 
is more easily seen in systems that develop inverse cascades, or that have 
non-local interactions \cite{bib:Mininni2011}

In this work, we consider time series stemming from experiments and direct
numerical simulations of hydrodynamic and MHD turbulence. Simulations are done
in periodic boxes, but with a volumetric forcing chosen to mimic the geometry
of the flow in the experiments. Parameters in the simulations are also chosen
to (within computational limitations) allow comparisons between the experiments
and the numerical results. Two different configurations are explored: the case 
with an externally imposed magnetic field (but with the magnetic field weak 
enough that can be considered a passive vector tracer and used to characterize 
the flow), and the case with self-sustained magnetic fields. The former case 
(corresponding to ``induction,'' as the magnetic field fluctuations observed 
are induced by motions of the conducting flow) can be relevant for many liquid 
metal laboratory experiments and for industrial flows. The latter, corresponding 
to a magnetohydrodynamic dynamo, has implications for magnetic field generation 
in geophysics and astrophysics.

\section{Experiments and simulations}

\subsection{Experiments}
\begin{figure}
\includegraphics[width=\columnwidth]{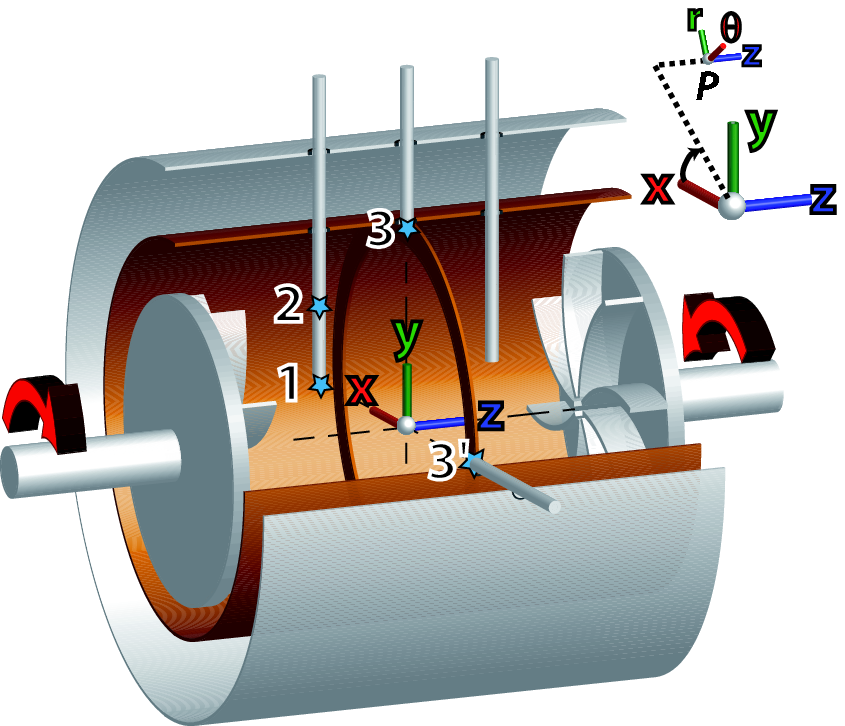}
\includegraphics[width=\columnwidth]{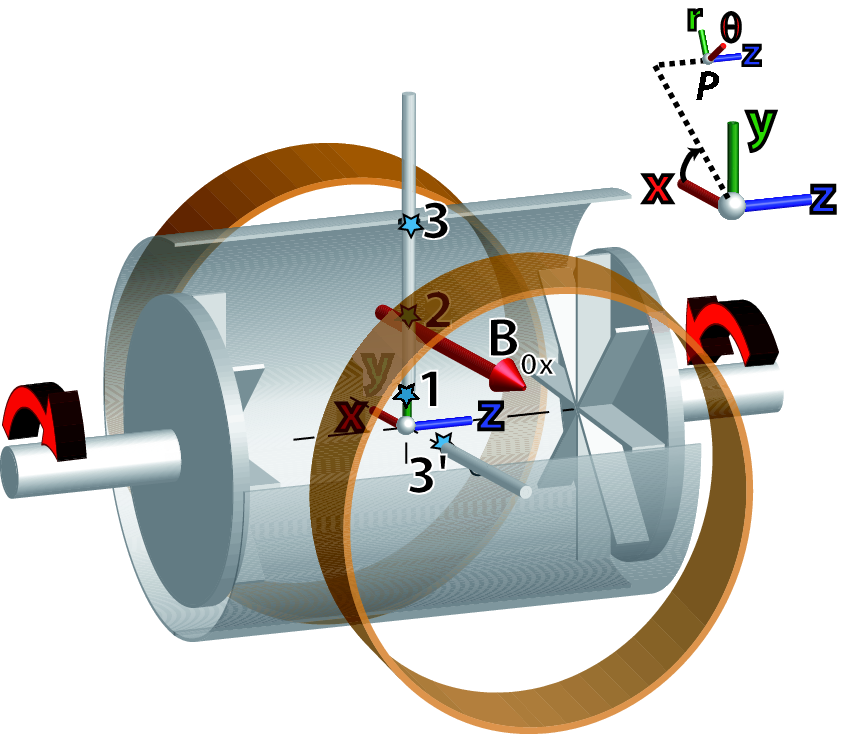}
\caption{({\it Color online}) Sketch of the experimental setups. 
  (a) VKS experiment drives 
  a von K\'arm\'an swirling flow of liquid sodium in a cylindrical
  vessel. Data from the experiments reported here was obtained in a
  configuration with an inner coper shell and an annulus in the
  mid-plane. (b) VKG experiment drives a von K\'arm\'an flow of liquid
  gallium in a cylindrical vessel. A pair of lateral Helmholtz coils
  allows us to apply a transverse magnetic field.
  Positions of the measurement probes are indicated with the stars.}
\label{fig:VKSetup}
\end{figure}

Experimental data discussed in this article has been obtained in the
VKG and VKS experimental setups (see Fig.~\ref{fig:VKSetup}). The
VKG experiment is a von K\'arm\'an swirling flow of liquid Gallium,
while VKS is a von K\'arm\'an flow of liquid Sodium. In both cases the
von K\'arm\'an flow is generated in the gap between two counter-rotating
impellers in a cylindrical vessel. This geometry has been extensively
investigated in the past decade in the context of magnetic field
self-generation by dynamo effect. Its interest in MHD studies has been
motivated by the large scale topology of the generated flow which
exhibits strong differential rotation and helicity.

The VKG experiment is mostly designed to investigate MHD mechanisms as
induction processes (transport and deformation of magnetic field lines
by the flow) and flow modification by an imposed magnetic field (via
the Lorentz force acting on the moving conducting fluid). The
reachable magnetic Reynolds number $Rm=UL/\eta$ (which compares the magnetic induction effects to magnetic dissipation effects, with $U$ the typical velocity of the flow, $L$ its typical dimension and $\eta$ the magnetic diffusivity)  accessible in VKG is of order unity
so that no dynamo effect is expected to be observed in this
experiment. On the other hand, the VKS facility has been designed and
optimized for dynamo generation. It is twice as large as VKG, with an
available mechanical power 15 times larger, and it uses liquid Sodium 
which is less resistive than Gallium. As a consequence, accessible 
magnetic Reynolds number in VKS is of the order of 50. In 2006 the 
first dynamo generation was observed in VKS when soft 
iron-impellers were used to drive the flow. Details of the VKG and VKS
experiments can be found in \cite{bib:monchaux2009_PoF,bib:volk2006_PoF,bib:verhille2010_SpaceSciRev}.

As for all liquid metals, the magnetic Prandtl number (defined as the
ratio of the kinematic viscosity $\nu$ of the fluid to its magnetic
diffusivity $\eta$, $\textrm{Pm}=\nu/\eta$) of Gallium and Sodium is very small
($\textrm{Pm}^{Ga} =1.4 \times 10^{-6}$, 
$\textrm{Pm}^{Na}=6.2 \times 10^{-6}$). As a consequence,
flows with magnetic Reynolds number $\textrm{Rm}$ of order unity or
larger have a kinematic Reynolds number 
$\textrm{Re}=UL/\nu=\textrm{Rm}/\textrm{Pm}$ exceeding $10^5$, hence
operating in highly turbulent conditions. Magnetic processes in VKG and VKS
experiments are therefore submitted to highly fluctuating turbulent
conditions. To this respect, von K\'arm\'an flows are particularly
interesting as they are known to present a wide hierarchy of scales of
fluctuations. At intermediate and small scales, the flow behaves as
\emph{traditional} turbulence (although anisotropic), with for
instance a classical $k^{-5/3}$ Kolmogorov spatial spectra. At larger
scales the flow undergoes in many cases long-term dynamics, which in
the absence of strong or self-generated magnetic fields, are mostly
driven by hydrodynamic instabilities of the mid-plane shear layer 
\cite{bib:crespoDelArco2009_GAFD,bib:caballero2013}.

As already mentioned, the VKG experiment is meant to investigate
magnetohydrodynamic induction processes with magnetic Reynolds number
$\textrm{Rm}<1$. To this purpose a pair  of transverse coils is placed
laterally to the experiment and allows to apply a transverse magnetic
field $B_{0x}$. When the applied magnetic field is sufficiently weak
so that the interaction parameter $N=\mu_0 L B_{0x}^2/\eta\rho U$ remains low and of the order of $10^{-3}$, the
imposed magnetic field can be considered as a passive vector,
advected by the flow. Recent experiments
\cite{bib:verhille2012_JFM} have shown that this approximation
stops to be valid as soon as $N \geqslant N^* \sim 0.02$, in which
case the flow itself is modified by the applied magnetic field. Only
the situation in which the magnetic field is weak and acts as a
passive vector  will be considered here. In this case, the magnetic 
field can just be used as a way to measure and characterize the 
hydrodynamic flow.

The VKS experiment has several possible configurations. We will focus
here on the geometry sketched in Fig.~\ref{fig:VKSetup} (top), which
has been extensively investigated and for which the main results have 
been reported in Ref.~\cite{bib:monchaux2009_PoF}. The flow is 
driven with iron impellers, such that a self-sustained dynamo 
magnetic field is observed above a magnetic Reynolds number 
threshold of the order $\textrm{Rm}^c\approx 30$.

Accessible measurements in VKS and VKG are mainly magnetic, as
well-resolved velocimetry is extremely difficult in liquid metals,
though some kinematic measurements have been recently made accessible
thanks to a local miniature Viv\`es probe~\cite{bib:noskov2009_PoF}. The induced or self-sustained magnetic field is
measured inside the flow by means of Hall probes. In the VKG
experiments, measurements were accessible on a line along the vertical
$Oy$ axis in the mid-plane (see the numbered stars in the 
bottom panel of Fig.~\ref{fig:VKSetup});
few measurements along the $Ox$ axis in the mid-plane are also
available. In the VKS experiments, measurements were accessible at the
locations indicated by stars in Fig.~\ref{fig:VKSetup} (top).

\subsection{Numerical simulations}
The numerical simulations solve the equations for an incompressible 
flow in a three-dimensional periodic domain of side $2\pi$. In the 
most general case (a conducting fluid with an externally imposed 
magnetic field of strength ${\bf B}_0$), the equations in
dimensionless Alfvenic units read
\begin{equation}
\frac{\partial {\bf u}}{\partial t} + {\bf u} \cdot \nabla {\bf u} =
     -\nabla {\cal P} + {\bf j} \times \left({\bf b}+{\bf B}_0 \right) +
      \nu \nabla^2 {\bf u} + {\bf F } ,
\label{eq:momentum}
\end{equation}
\begin{equation}
\frac{\partial {\bf b}}{\partial t} + {\bf u} \cdot \nabla {\bf b} =
      \left({\bf b}+{\bf B}_0 \right) \cdot \nabla {\bf b} +
      \eta \nabla^2 {\bf b} .
\label{eq:induction}
\end{equation}
In the incompressible case $\nabla \cdot {\bf u} = 0$, and
$\nabla \cdot {\bf b} = 0$. Here, ${\bf u}$ is the velocity, ${\bf b}$ is
the magnetic field, and ${\bf j}=\nabla \times {\bf b}$ is the current
density. The pressure (normalized by the density) is ${\cal P}$,
$\nu=4.7 \times 10^{-3}$ is the dimensionless kinematic viscosity, 
and $\eta$ the magnetic diffusivity, with $\eta = \nu/\textrm{Pm}$.
In the absence of an externally imposed magnetic field 
(${\bf B}_0=0$) these equations are the incompressible MHD equations
often used to study the dynamo effect. When ${\bf B}_0={\bf b}=0$ (or when 
the interaction parameter $N$ is small enough), Eq.~(\ref{eq:momentum}) 
reduces to the Navier-Stokes equation for an incompressible hydrodynamic 
flow.

\begin{figure}
\includegraphics[width=8.5cm]{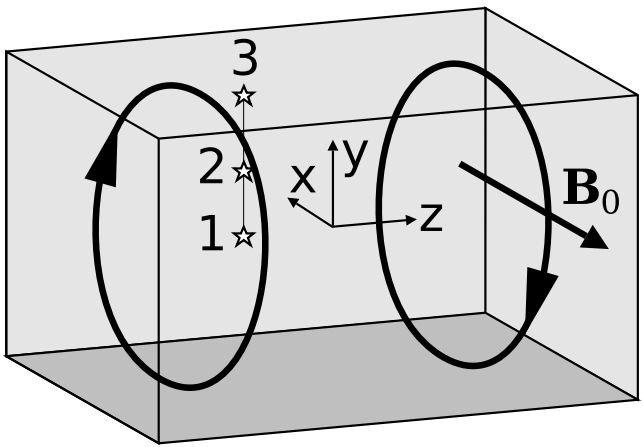}
\caption{Sketch of the numerical Taylor-Green cell, and position of the
points where magnetic and velocity measurements are made (indicated
with stars). The two large circular arrows represent the two
counter-rotating Taylor-Green vortices.}
\label{fig:geometry}
\end{figure}

To mimic the geometry of the flow in the experiment, a Taylor-Green (TG)
vortex is used as mechanical forcing
\begin{equation}
{\bf F} = A_0 \left[ \begin{array}{c}
     \sin(k_0 x) \cos(k_0 y) \cos(k_0 z)  \\
     -\cos(k_0 x) \sin(k_0 y) \cos(k_0 z) \\
     0  \end{array} \right] ,
\label{eq:TG}
\end{equation}
where $A_0=0.41$ is the forcing amplitude, and $k_0=2$ is the forcing
wavenumber. This forcing was shown before to give good agreement with
von K\'arm\'an dynamo results \cite{bib:ponty2005_PRL,bib:Mininni2005_AJ},
while allowing simulations to be done in periodic boundaries and thus
lowering computational costs. A sketch of the resulting flow in one TG
cell can be seen in Fig.~\ref{fig:geometry}.

Equations (\ref{eq:momentum}) and (\ref{eq:induction}) are solved
pseudospectrally using a parallelized code
\cite{bib:gomez2005_ScriptA,bib:gomez2005,bib:hybrid2011} 
dealiased with the $2/3$-rule. Second-order Runge-Kutta 
is used to evolve the equations in time. Since long 
integrations are needed (in many cases simulations were continued
for over 6000 turnover times), linear spatial resolution is $N_{res}=128$ in all
cases. To compare with experiments, time series of the three Cartesian
components of the velocity and the magnetic fields at the three points
indicated in Fig.~\ref{fig:geometry} are recorded with high cadence. Also,
time series of the amplitude and phase of each Fourier mode in spectral
space are stored, to be able to identify the modes responsible for long-term 
behavior. Note that with the spatial resolution considered here, magnetic 
Reynolds numbers similar to the ones in the experiments are easily 
reproduced. However, the mechanical Reynolds numbers in the experiments are 
out of reach for even the largest resolutions we could attain in simulations, 
and those resolutions would preclude the long time integrations considered 
in this study.

\begin{figure}
\includegraphics[width=\linewidth]{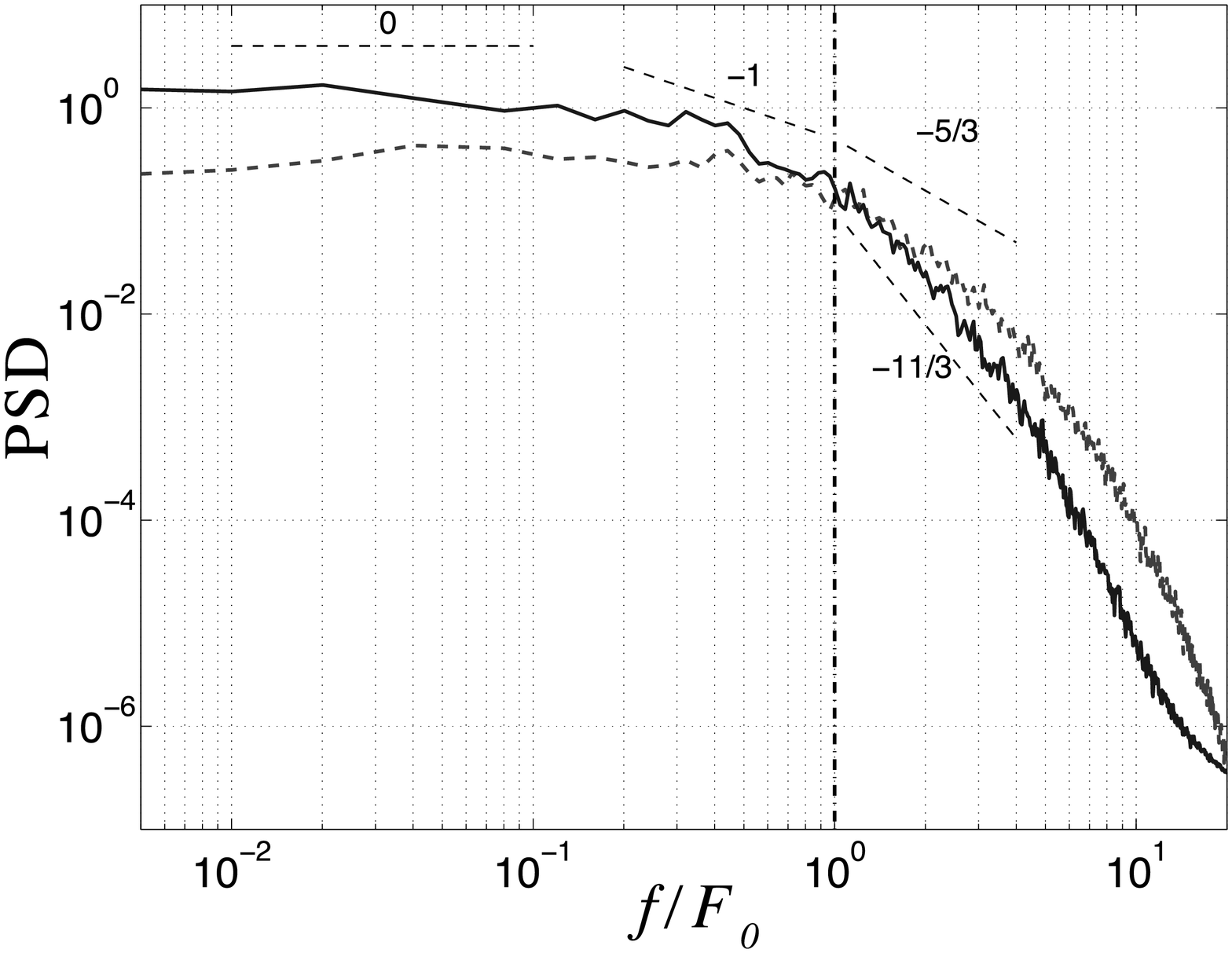}
\includegraphics[width=\linewidth]{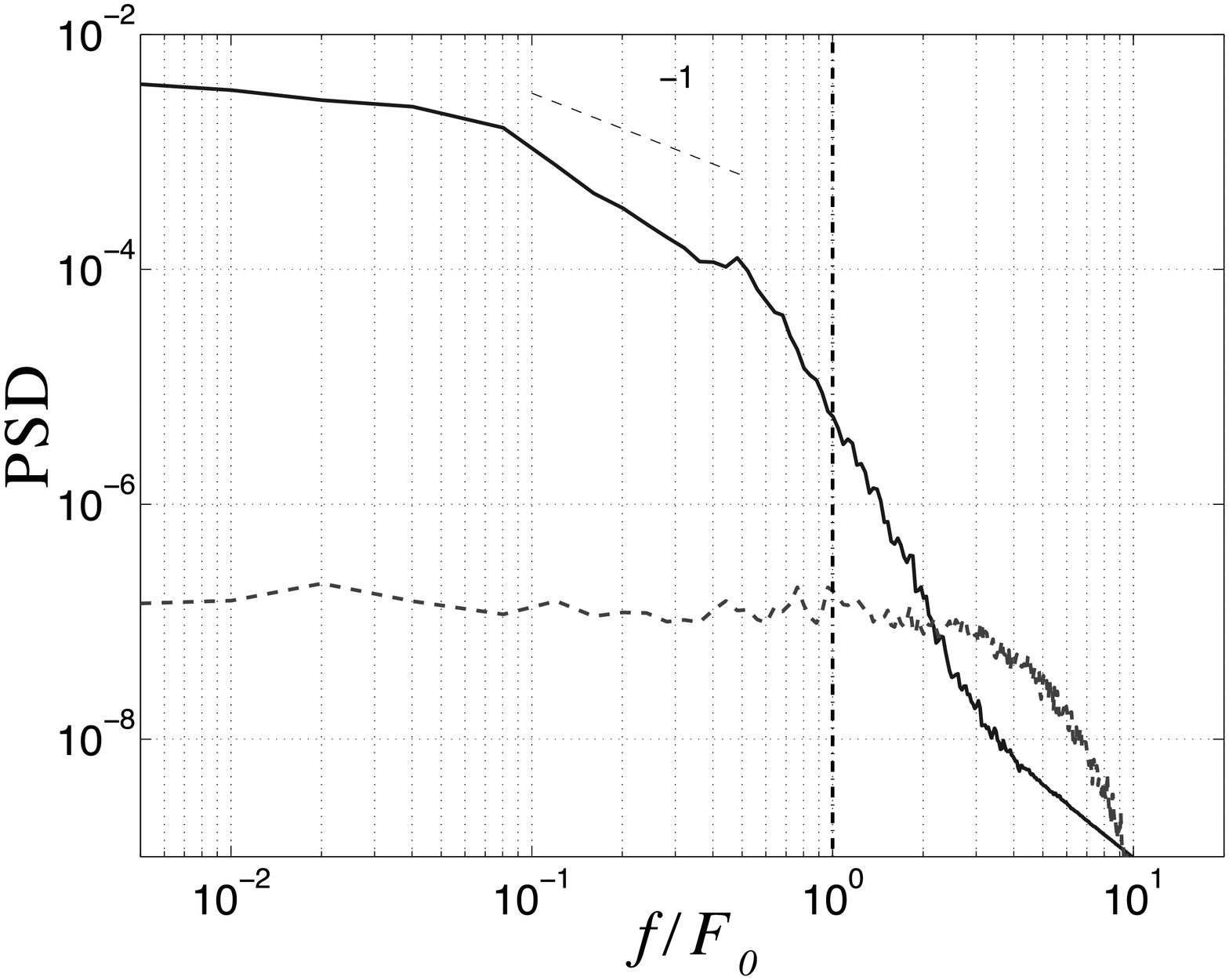}

\caption{Numerical simulations of Taylor-Green MHD induction. Top: Power spectrum of the temporal fluctuations of the axial
  velocity $u_z$ at point 1 (light gray dashed line) and of the axial
  induced magnetic field (dark gray solid line) in the numerical
  simulations at low magnetic Prandtl number ($\textrm{Pm}=0.05$ and low
  interaction parameter $N=4\times 10^{-5}$. Frequency axis has been
  normalized by the forcing frequency $F_0=u_{rms}/\pi$. The global
  energy level of the spectra has been arbitrarily set to make them
  coincide for $f=F_0$. Bottom: Power spectrum of the amplitude of
  Fourier mode $k=(1,0,0)$ of $u_z$ (dark gray solid line), and of 
  Fourier mode $k=(10,10,10)$ of $u_z$ (light gray dashed line). In
  both panels, several slopes reported before in the VKS and VKG
  experiments are shown as references, although we are mostly
  interested in the behavior for $f<F_0$.}
\label{fig:hdtimeseries}
\end{figure}

The strategy used in the simulations is as follows. The mechanical force is
turned on at $t=0$ from the flow at rest, and the Navier-Stokes equations
(with the ${\bf b}$ and ${\bf B}_0$ fields set to zero) are advanced for over
6000 turnover times (a turbulent steady state is reached shortly after 10
turnover times). Time series of the velocity are then used for the analysis of
$1/f$ noise in hydrodynamic flows (as well as to compare with the equally long 
simulations with a weak imposed magnetic field). Later, the last state of the 
hydrodynamic flow is used to start two different sets of simulations: one in 
which a uniform magnetic field ${\bf B}_0=B_0 \hat{y}$ is imposed, and one 
in which ${\bf B}_0=0$ and an initially small magnetic seed is amplified 
until an MHD steady regime is reached. The former is intended to mimic 
induction experiments with a transverse magnetic field in VKG, while the latter 
is intended to mimic dynamo experiments in VKS. In both cases the
systems are integrated for over $6000$ turnover times. The magnetic 
Prandtl number in induction simulations is $\textrm{Pm} = 0.05$ 
and the interaction parameter is adjusted by setting $B_0$ such 
that $N \approx 10^{-4}$. In dynamo simulations 
$\textrm{Pm} = 0.5$ as a larger magnetic Reynolds number is 
needed to sustain magnetic fields 
\cite{bib:ponty2005_PRL,bib:Mininni2005_AJ}. The mechanical Reynolds
number is fixed in all simulations, and is $\textrm{Re} \approx 670$.

\begin{figure}
\includegraphics[width=\linewidth]{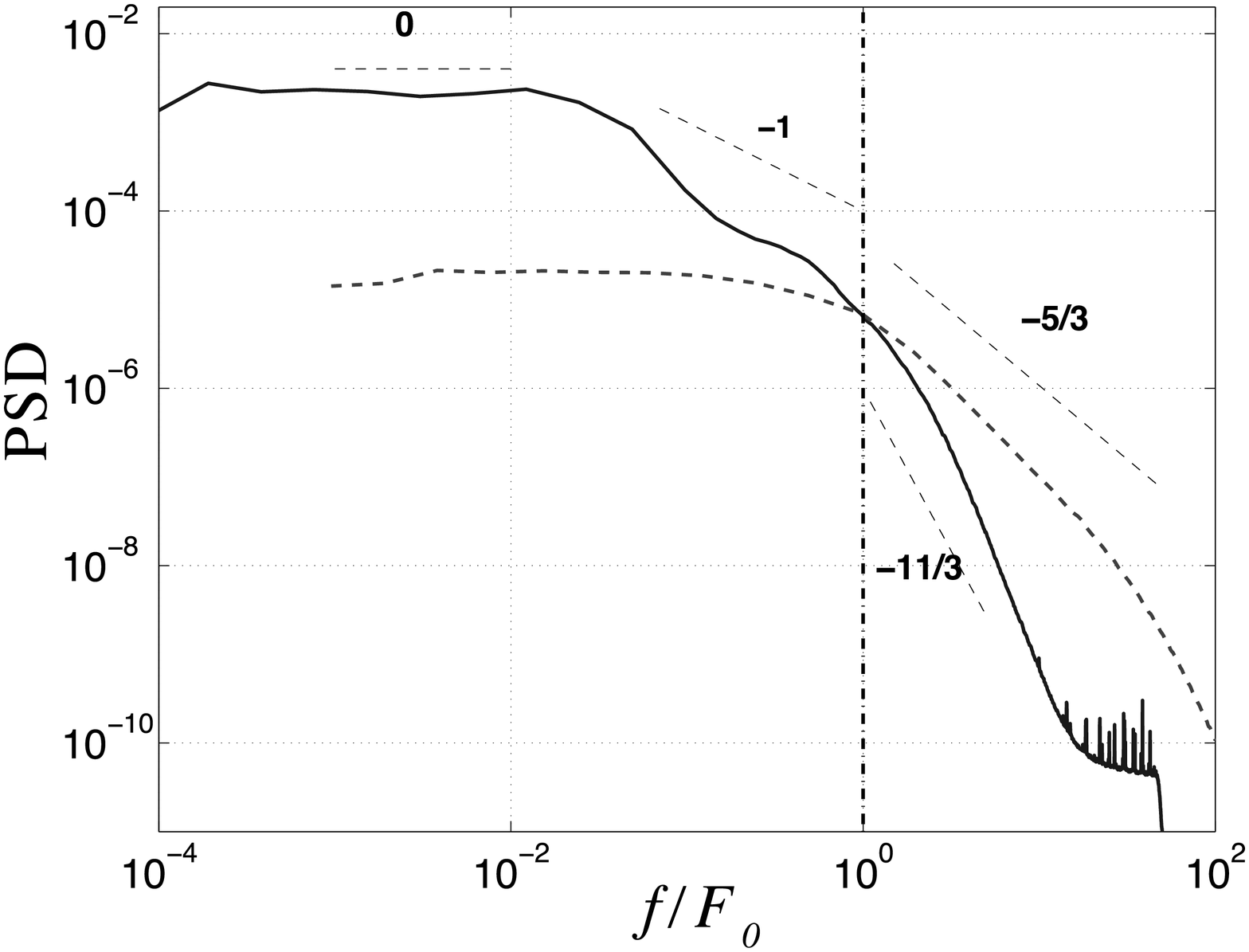}
\caption{Experiments of MHD induction in VKG flow. Power spectrum of the axial velocity (light gray
  dashed line) and of the induced axial magnetic field in VKG
  experiment with a weak imposed transverse magnetic field. Frequency
  axis has been normalized by the forcing frequency $F_0$ taken as the
  rotation rate of the impellers $\Omega$ (10~Hz for the shown
  experiment). The global energy level of the spectra has been
  arbitrarily set to make them coincide for $f=F_0$. Several slopes 
  are shown as references.}
\label{fig:HydroExpBvsV}
\end{figure}

\begin{figure}
\includegraphics[width=\linewidth]{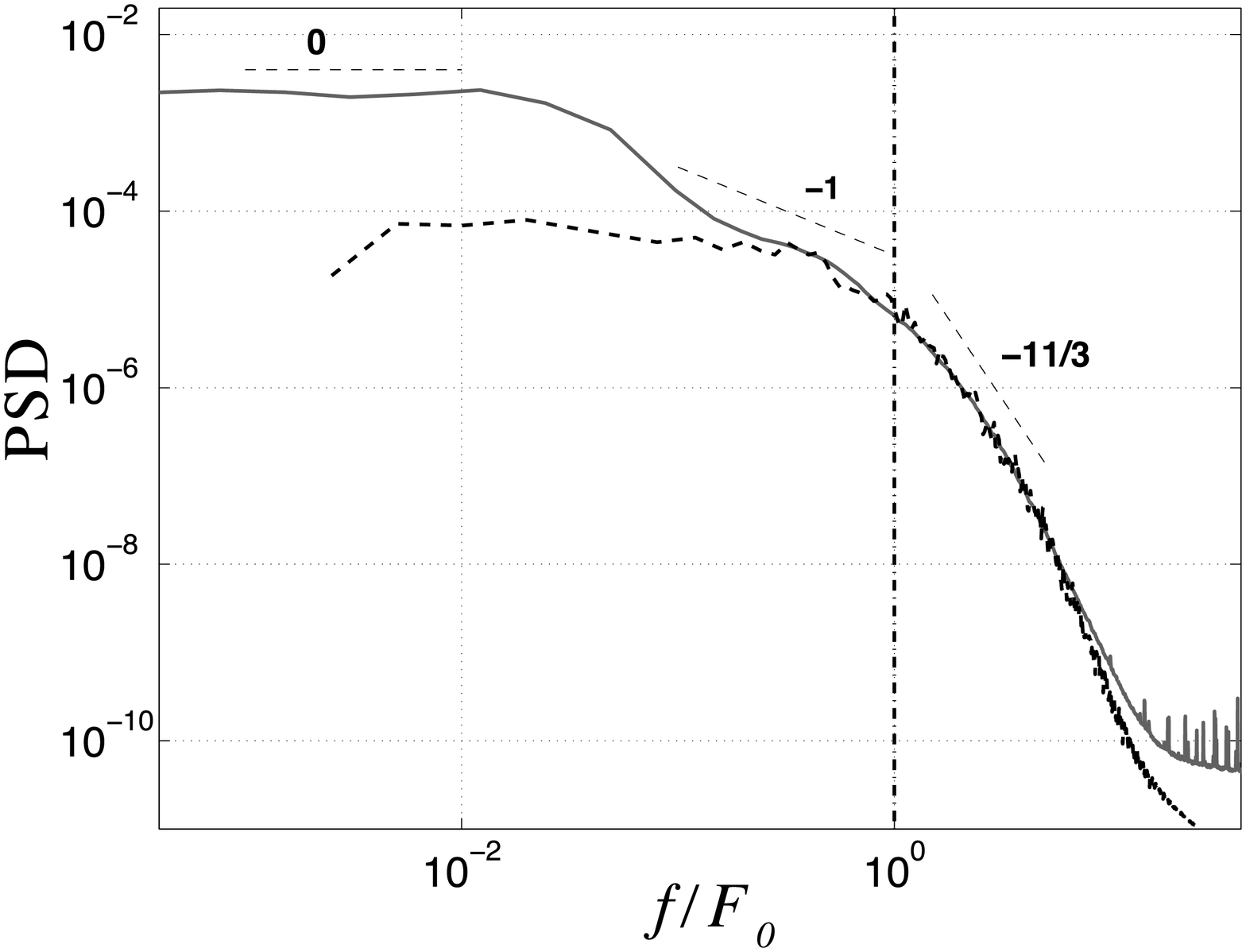}
\caption{Experiments vs. simulations (induction case): comparison of the spectra of the 
  induced magnetic field in figures \ref{fig:hdtimeseries}
  (simulation) and \ref{fig:HydroExpBvsV} (experiment). The total
  energy level has been arbitrarily set to make both spectra coincide
  at $f=F_0$. Slopes are shown as references. Measurements from the 
  simulation and the experiment agree remarkably well except for very 
  low frequencies.}
\label{fig:HydroNumvsExp}
\end{figure}

\section{Hydrodynamic turbulence and low $N$ induction}

When the imposed magnetic field is weak enough that it can be 
considered a passive vector, time series of the velocity field
measured in the experiments and simulations at any of the points 
indicated in Figs.~\ref{fig:VKSetup} and \ref{fig:geometry} do not 
show clear long-term behavior or $1/f$ noise, as it can be seen for 
instance from the dashed lines in Fig.~\ref{fig:hdtimeseries} (top) 
and \ref{fig:HydroExpBvsV}, which represent the kinetic spectra 
for the simulation and the experiment respectively. A relatively 
clear inertial range of time scales, consistent with a classical 
$f^{-5/3}$ Kolmogorov spectrum is present, but no significant long 
term dynamics is observed at low frequencies, and the spectra 
flatten rapidly below the forcing frequency $F_0$ (in the experiment 
$F_0$ is taken as $\Omega$, the rotation rate of the impellers, while 
in the simulation $F_0$ is determined as $u_{rms}/L$, where 
$L=2\pi/k_0 = \pi$ is the forcing scale and $u_{rms}$ is the 
root mean square of the turbulent velocity fluctuations).

The absence of long-term dynamics is in agreement with previous 
results indicating isotropic and homogeneous hydrodynamic 
turbulence does not display $1/f$ noise \cite{bib:dmitruk2007_PRE}. 
However, and unlike the results in Ref.~\cite{bib:dmitruk2007_PRE},
when in the numerical simulations the time series of the amplitude 
of individual Fourier modes are considered, we find that modes in 
the $k=1$ shell do have $1/f$ spectra (the observed spectrum is 
actually steeper than $1/f$, see the bottom panel of 
Fig.~\ref{fig:hdtimeseries}). Interestingly, it can be
noted that the modes in this Fourier shell break down the symmetries 
of the TG flow and, as an example, are responsible for large-scale 
fluctuations in the $z$ position of the shear-layer between the two 
Taylor-Green vortices. Indeed, $1/f$ spectra are only observed in modes 
that break down this symmetry. The spectrum of the temporal 
fluctuations of small-scale modes (large wave-numbers) does not 
exhibit any long-term dynamics, as it can be seen in 
Fig.~\ref{fig:hdtimeseries} (bottom).  However, at least at the Reynolds 
number considered in the simulation, the modes that break down 
the symmetry do not have enough energy to give rise to a clear 
long-term behavior when all Fourier modes are integrated to obtain 
the velocity at one of the measurement points (as shown in the top 
panel of Fig.~\ref{fig:hdtimeseries}).

As already mentioned, when modes with $1/f$ behavior are not 
the most energetic, $1/f$ behavior is not observed in global 
quantities. However, as soon as a physical effect makes these 
modes dominant, $1/f$ noise arises. This can be observed for 
instance by considering the evolution of the small imposed 
magnetic field with low magnetic Prandtl number and low 
interaction parameter ($N \ll 1$). Condition $N \ll 1$ ensures 
that the magnetic field passively traces the carrier velocity field, 
while condition $\textrm{Pm}\ll 1$ imposes a large scale separation 
between magnetic and velocity field, so that only large scales 
of the velocity field are traced by the magnetic field. Figure 
\ref{fig:hdtimeseries} (top) also shows the spectrum of the axial 
induced magnetic field, in the presence of a small applied transverse 
field, from the numerical simulation (with $\textrm{Pm}=0.05$ and 
$N=4 \times 10^{-5}$). A short but still visible $1/f$ regime 
appears for frequencies below the forcing frequency $F_0$. 
As previously noted, this regime is not visible for the velocity 
spectrum. For the comparison, the global energy level of the kinetic
and magnetic spectra has been set so as to make them coincide at
$f=F_0$.

\begin{figure}
\includegraphics[width=\linewidth]{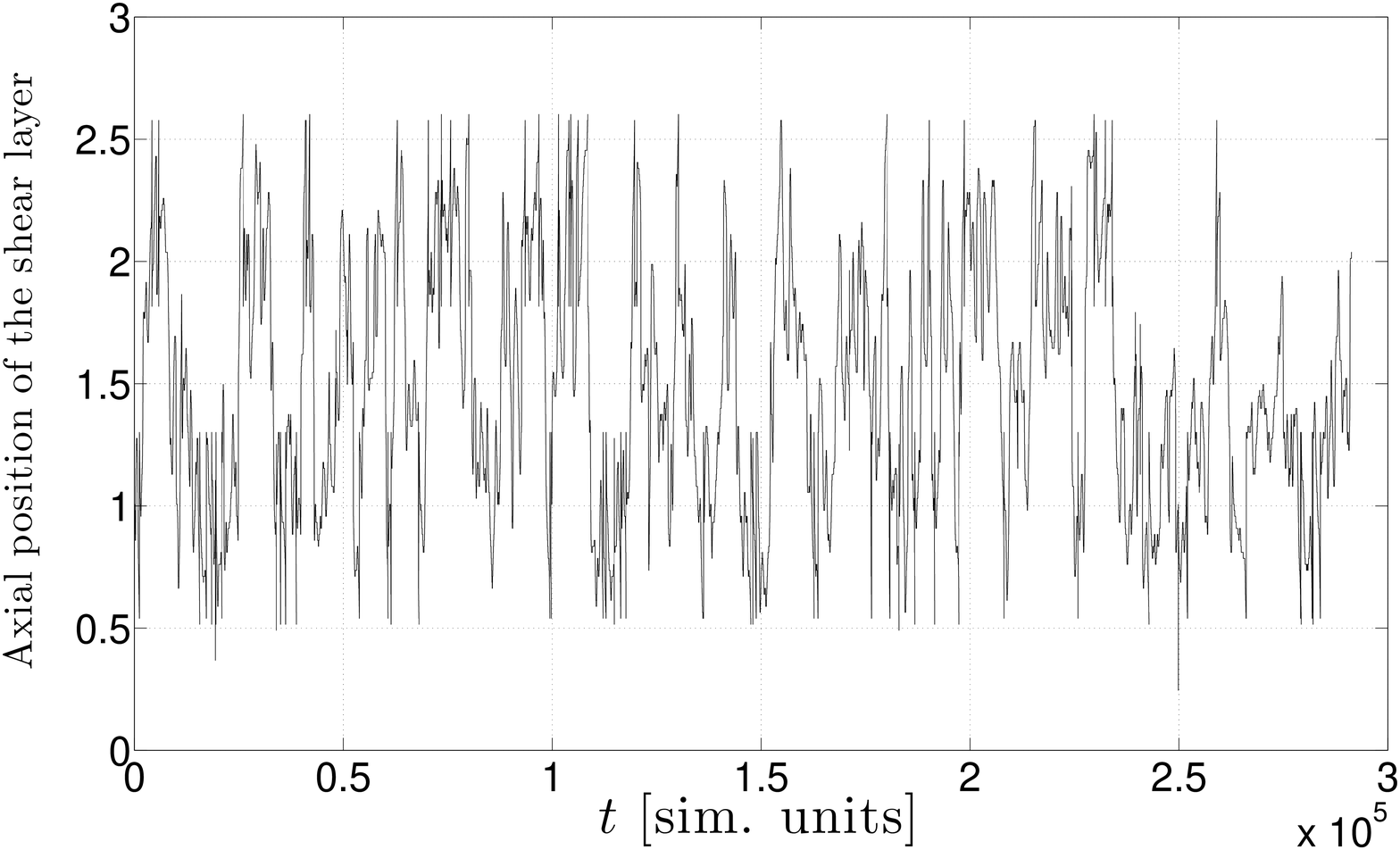}
\includegraphics[width=\linewidth]{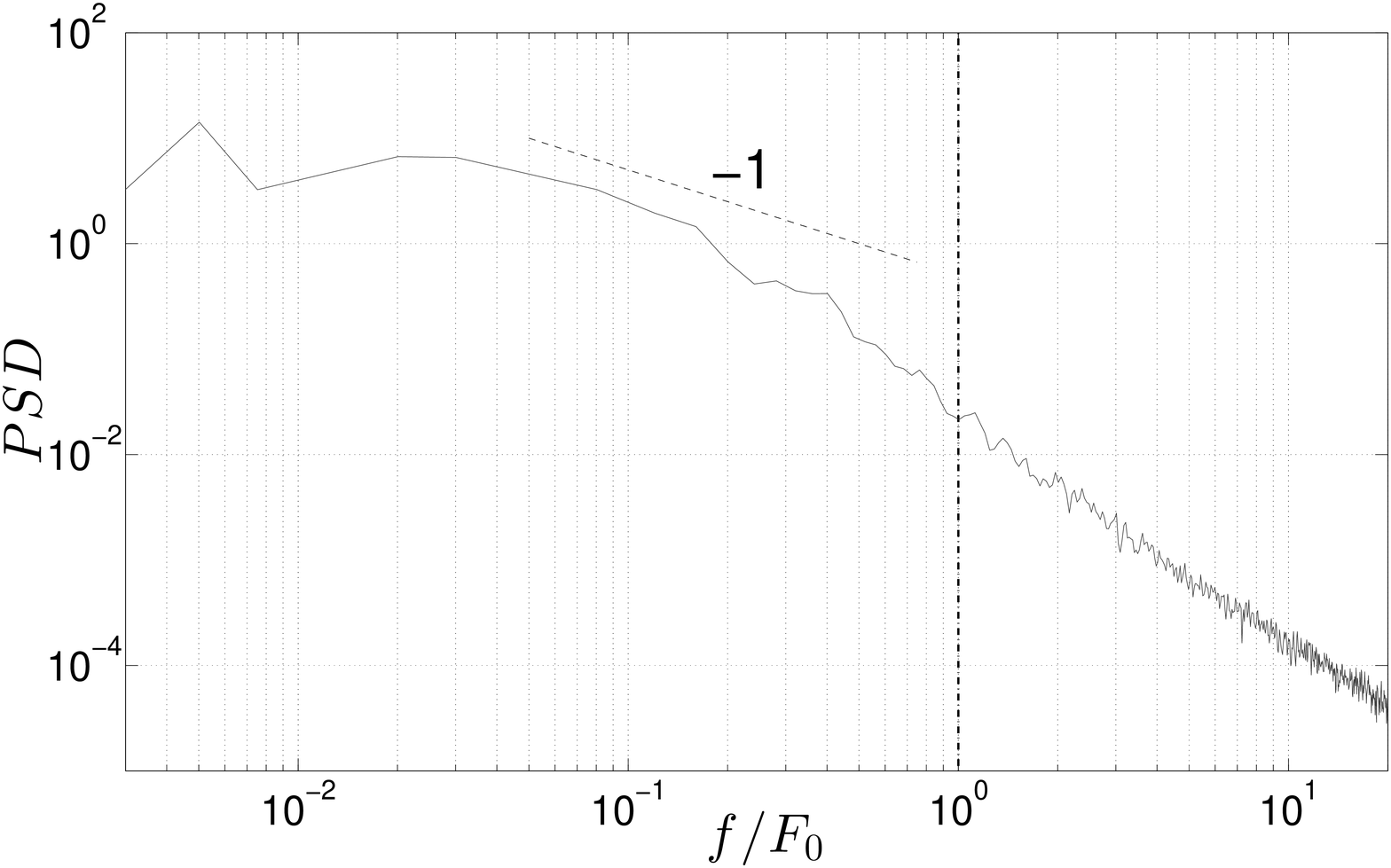}
\caption{Numerical simulations. Top: position of the shear layer as a function of time (in simulation units). Bottom: power spectrum of the signal in the top
figure. A slope of $-1$ is indicated as a reference.}
\label{fig:numshearlayer}
\end{figure}

\begin{figure}
\includegraphics[width=\linewidth]{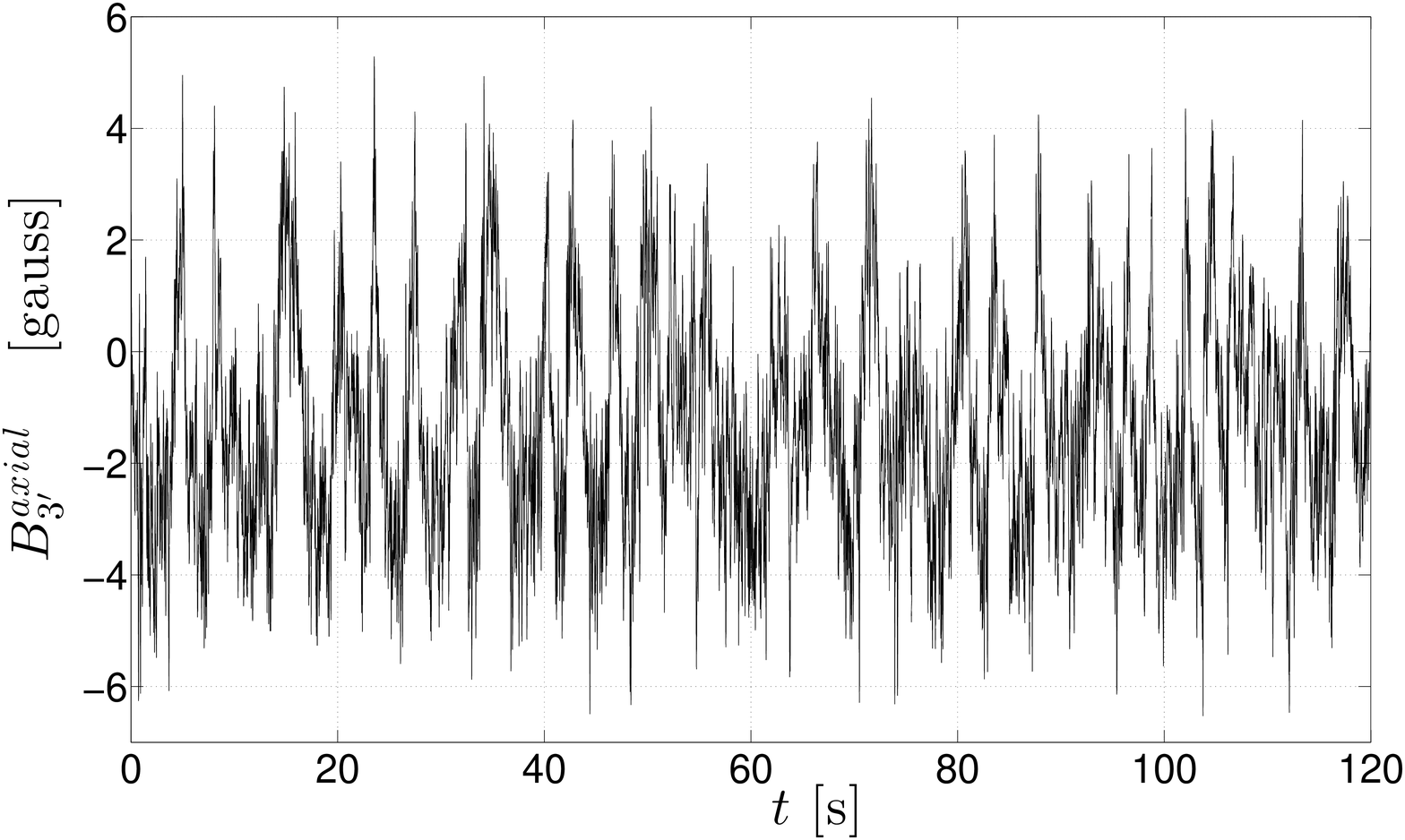}
\includegraphics[width=\linewidth]{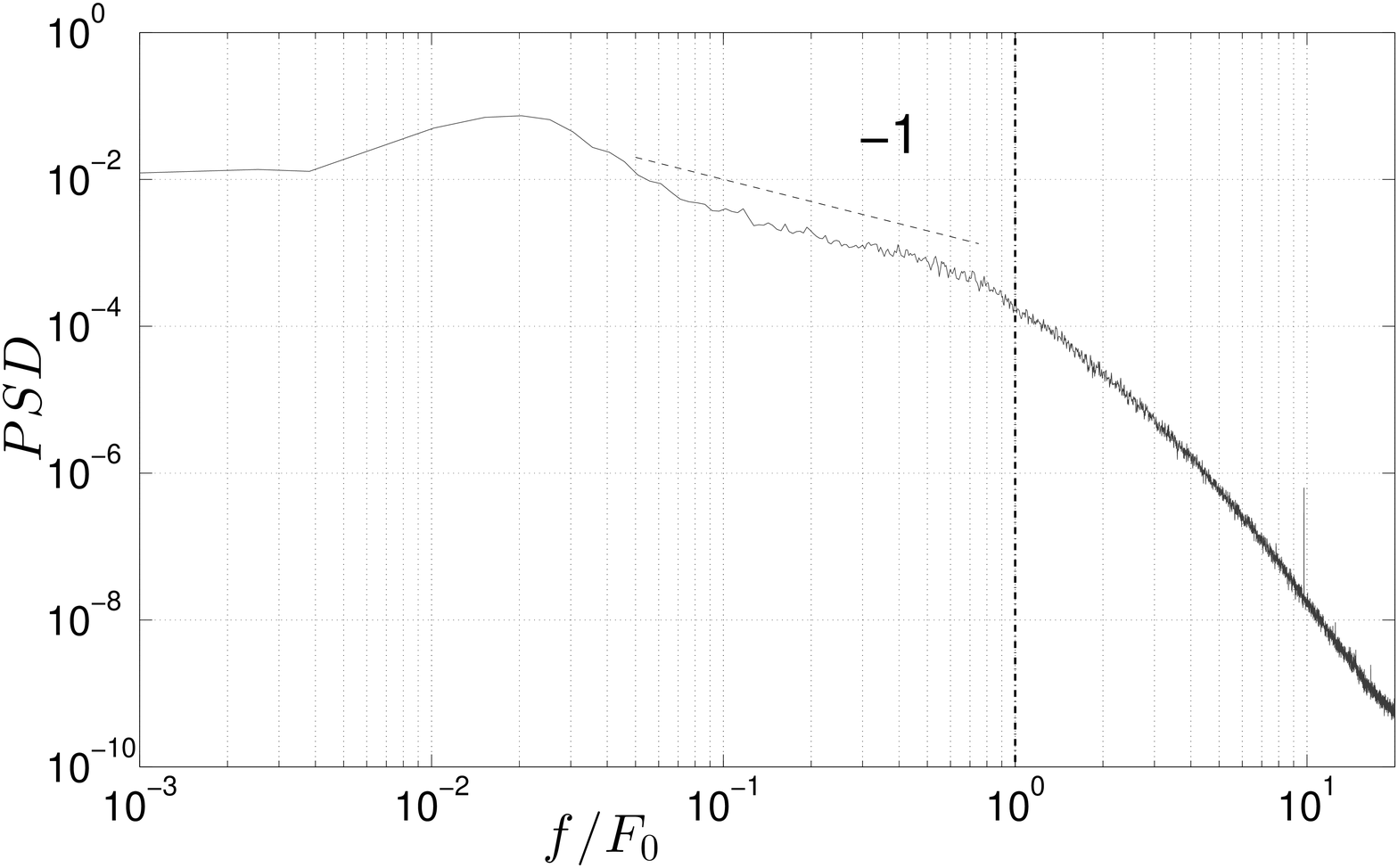}
\caption{VKG experiment. Top: proxy of the evolution of the shear layer as a function of
time, as indicated by the evolution of the axial induced
magnetic field measured at point 3'. By ${\cal{R_\pi}}$ symmetry
around the $Ox$ axis, this component of induced field must be
zero. However, it is visible from the measurements that it fluctuates
between a high level state (of order 4 gauss) and a low level state
(of order -4 gauss). We interpret these fluctuations as a signature of
the instantaneous breaking of the ${\cal{R_\pi}}$ symmetry due to the
motion of the shear layer. Bottom: power spectrum of the signal in
the top figure. A slope of $-1$ is indicated as a reference.}
\label{fig:expshearlayer}
\end{figure}

Similarly, Fig.~\ref{fig:HydroExpBvsV} shows the spectrum of the
velocity measured in VKG experiment with a Viv\`es probe compared to
the spectrum of the induced axial magnetic field (measured at point 1)
when a weak passive transverse magnetic excitation $B_{0x}$ is imposed
(note that high frequency resolution of the Viv\`es probe is limited
by magnetic diffusion at the probe size, which is responsible for the
cut-off observed here for frequencies $f\gtrsim200$~Hz~\cite{bib:noskov2009_PoF}). For the
comparison, the global energy level of the kinetic and magnetic
spectra has been set so as to make them coincide at $f=F_0$. Contrary
to the kinetic spectrum, the magnetic one shows an important long term
dynamics below $F_0$, that, at the light of the previous numerical
observations, shall be attributed to the large scale modes of the
kinetic field naturally traced by the magnetic field in such a low 
magnetic Prandtl number regime. In other words, the passive 
magnetic field at very low $\textrm{Pm}$ acts as a low-pass 
filter that only senses the velocity field at very large scales.

\begin{figure*}
\includegraphics[height=8.2cm]{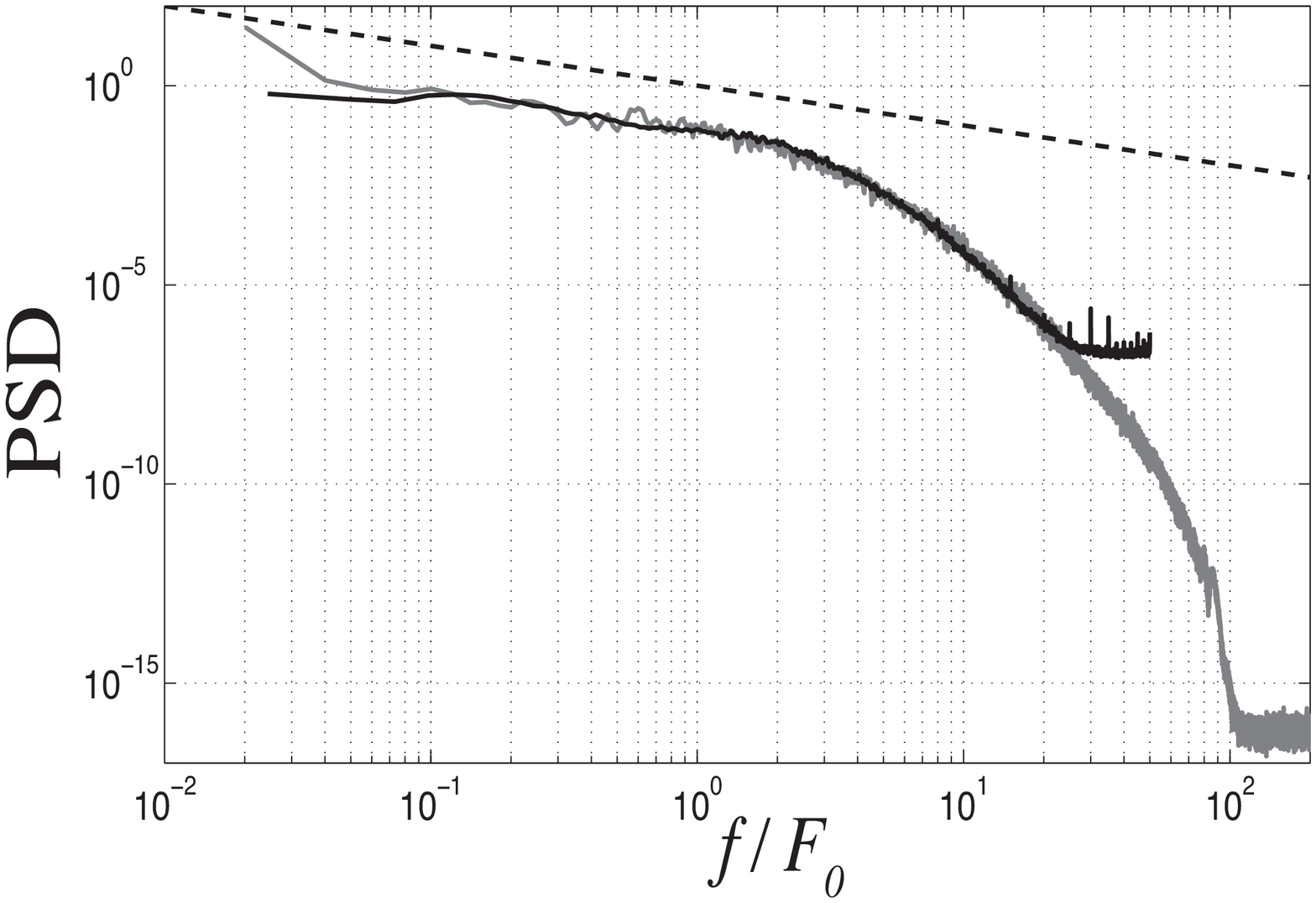}\\
\includegraphics[height=4.5cm]{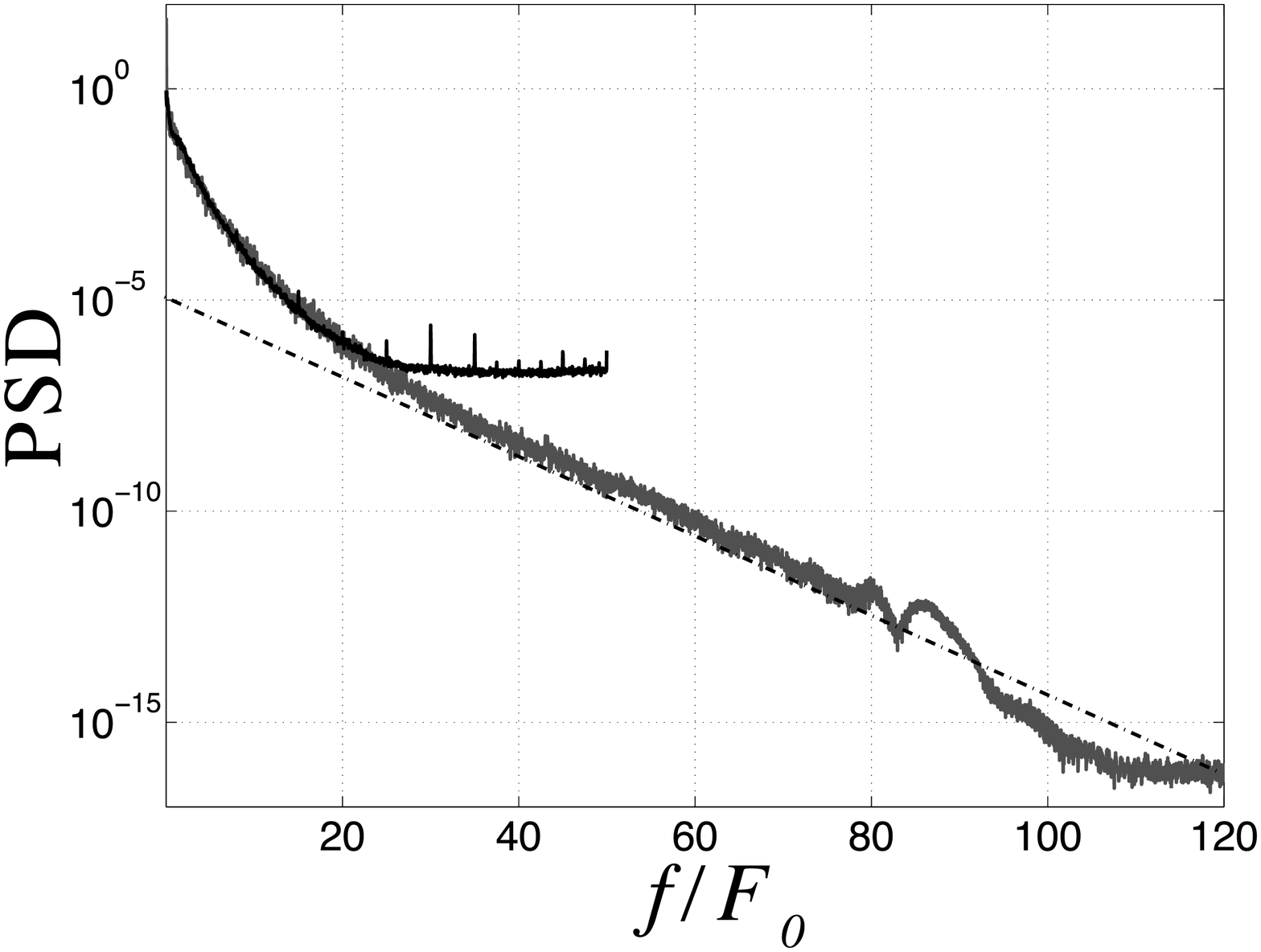} 
\includegraphics[height=4.5cm]{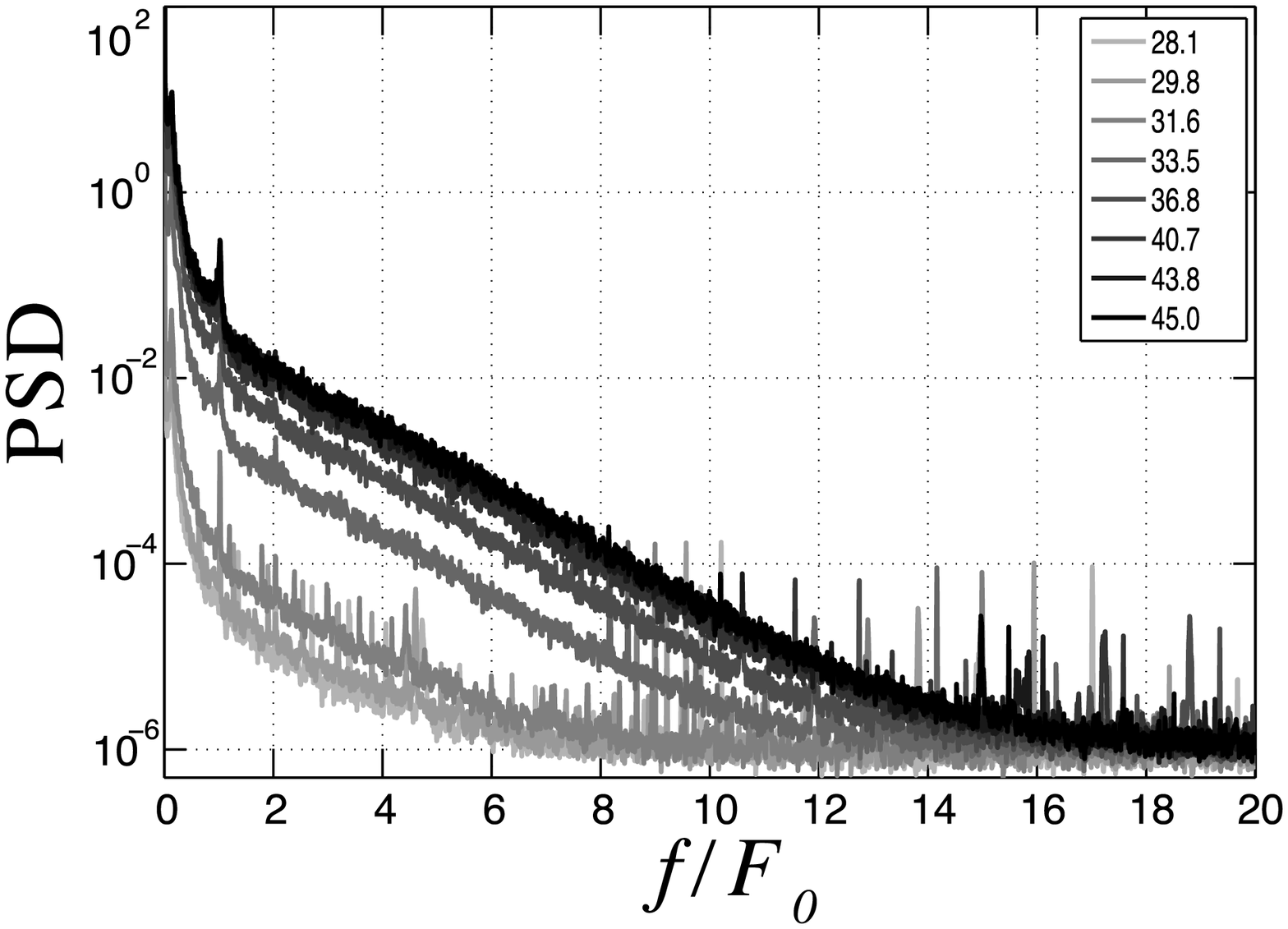}
\caption{Experiments vs. simulations in the dynamo case. Top: power spectra of the self sustained magnetic field in 
the VKS experiment (black) and in the simulation (gray). A slope of 
$-1$ is indicated as a reference. Bottom left: semilog plot of top 
figure. Bottom right: semilog plot of magnetic spectra measured at 
point 3 in VKS (from Ref.~\cite{bib:monchaux2009_PoF}). The different 
curves correspond to different magnetic Reynolds numbers as 
indicated in the legend, with increasing $\textrm{Rm}$ from light 
gray to black curves.}
\label{fig:dyncomparison}
\end{figure*}

To push further the comparison between the experiment and the
simulation, Fig.~\ref{fig:HydroNumvsExp} shows the magnetic spectra
from the simulation and the experiment in Figs.~\ref{fig:hdtimeseries} 
and \ref{fig:HydroExpBvsV} ($axial$ component
of the induced magnetic field measured at point 1 in the simulation
and at point 1 in VKG, which corresponds to the closest available
geometries between experiments and simulations). The total
energies have been rescaled so that the spectra in the numerics and
in the experiment are equal for $f=F_0$. We recall that the frequency
axis has been normalized by the corresponding forcing frequency
$F_0$: $F_0=u_{rms}/\pi$ for the simulation, and $F_0$=$\Omega$ the
rotation rate of the disks for the experiment. Interestingly, this
scaling was found to give the best collapse between numerics and
experiment. As it can be seen in the figure, the agreement is indeed
remarkable, except for the very low frequencies which are enhanced
in the experiment. The slow $1/f$ dynamics for frequencies
$f\lesssim F_0$ is in particular observed in both studies. It is also
interesting to note that this agreement for the dynamical features of
the induction processes is observed even if the dominant induction
mechanisms are different in the numerics and the experiment. It has
indeed been shown in the experiment that the \emph{mean} axial
component induced at the location of the measurement point 1 when a
transverse field $B_{0x}$ is applied results from a boundary condition
effect due to conductivity difference between the wall of the
experiment and the fluid, combined to the presence of local flow
vorticity in the vicinity of the boundary \cite{bib:bourgoin2004_PoF}
(BC mechanism). Such a mechanism cannot be present in the numerics
where the entire medium has a homogenous conductivity. This may
explain the difference between the simulation and the experiment 
at very low frequencies, which in the experiment are dominated by 
this very specific average induction processes.

However, in both cases the fluctuations of induction processes 
do trace back the fluctuations of the flow. The long-term effects 
at $f\lesssim F_0$ for intermediate frequencies must therefore 
be associated with fluctuations in the velocity field that are 
present in both the experiments and simulations. In other 
words, the remarkable collapse of numerical 
and experimental spectra for moderate and high frequencies 
indicates that the magnetic field fluctuations at these 
frequencies are tracing fluctuations of the velocity field with 
a common origin.

The numerical observation that $1/f$ is essentially related to large
scale hydrodynamic modes breaking the symmetries of the TG flow, and
the experimental observation of a $1/f$ regime in a configuration
where one of the sources of induction is related to radial vorticity in
the mid-plane, both point toward an important role (for intermediate 
frequencies) of the fluctuations of the strong shear layer in the 
mid-plane of the von K\'arm\'an flow. This shear layer is indeed 
known to undergo strong large scale and long term fluctuations in 
experimental von K\'arm\'an flows 
\cite{bib:crespoDelArco2009_GAFD,bib:caballero2013,bib:ravelet2008_JFM}. 
These fluctuations are also observed in the 
simulation, and are shown in Fig.~\ref{fig:numshearlayer}, 
which represents the time series of the
position of the shear layer as a function of time, and its power
spectrum in the numerics. To identify the position of the shear layer
in the simulations, a low-pass filtered and averaged (in $x$ and $y$) 
velocity field was computed in the TG cells from the Fourier 
coefficients of the velocity, leaving mean profiles for the 
azimutal and $z$ component of the velocity field that 
depend only on $z$ (see Fig.~\ref{fig:geometry}).

In the VKG experiments, a proxy of the evolution of the shear layer
can be obtained from measurements of magnetic induction. A
particularly clear tracing of the shear layer is obtained when
measuring the induced axial magnetic field $B^i_x$ at a location on
the horizontal $Ox$ axis (point 3' in Fig.~\ref{fig:VKSetup}) for an
applied transverse magnetic field $B_{0x}$. Simple symmetry
considerations impose indeed the induced axial magnetic field measured
at any point of the $Ox$ axis for an applied transverse field $B_{0x}$
to be zero on average \cite{bib:bourgoin2004_MHD}. Figure
\ref{fig:expshearlayer} (top) shows the corresponding time series
which clearly shows that the measured signal does fluctuate around
zero, but that it explores induction regimes with symmetric polarities
which can be attributed to slow symmetry breaking fluctuations of the
shear layer respective to its equilibrium position in the mid-plane
\cite{bib:bourgoin2004_MHD}. Figure \ref{fig:expshearlayer} (bottom)
shows the corresponding spectrum which does exhibit a clear long-term
$1/f$ regime.

\section{Magnetohydrodynamic dynamo}

\begin{figure}
\includegraphics[width=\linewidth]{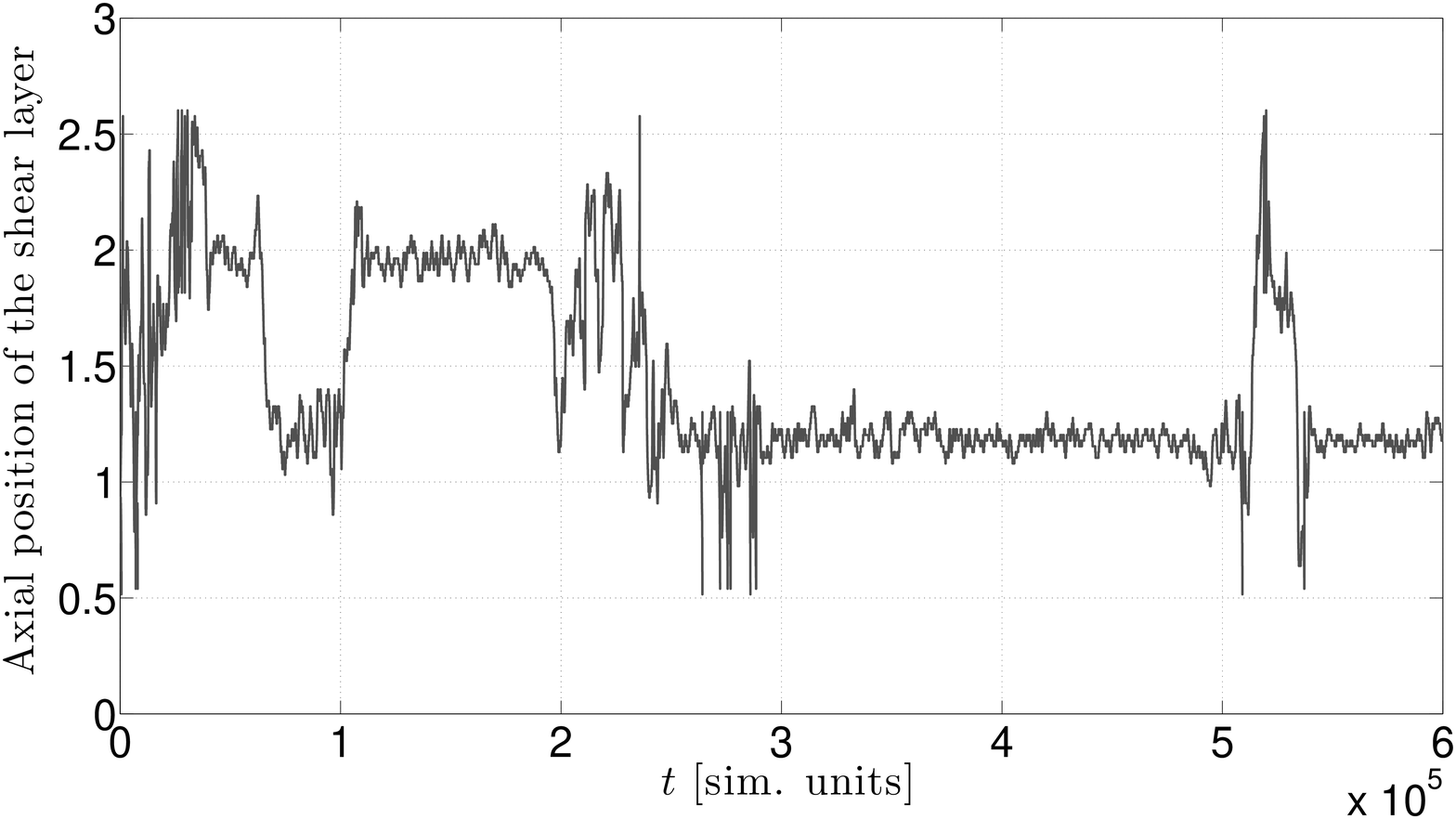}
\includegraphics[width=\linewidth]{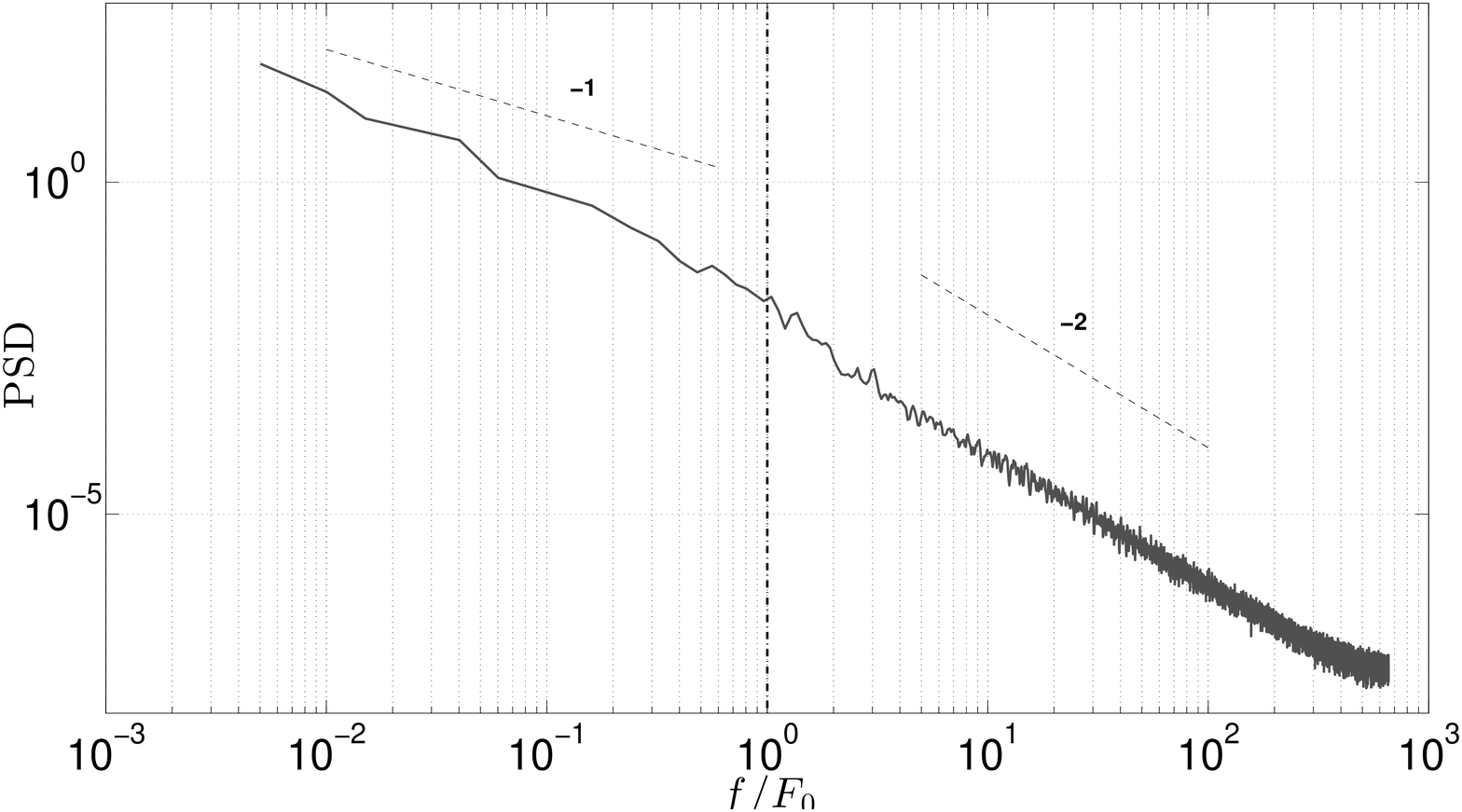}
\caption{Position of the shear layer as a function of time in the numerical
simulations of dynamo action, and power spectrum of the signal. A
slope of $-1$ is indicated as a reference.}
\label{fig:dynshearlayer}
\end{figure}

We now turn to the case of MHD dynamo and compare dynamical
fluctuations observed in the VKS experiment and numerical results from
the present TG simulations. Though parameters are not really
comparable (in particular the Prandtl number differs by orders of
magnitude as it is of the order of $6\times 10^{-6}$ in VKS while it is $0.5$
in the simulation), experimental and numerical spectra of magnetic
fluctuations show a remarkable good agreement, with in particular a
wide $1/f$ range as shown in Fig.~\ref{fig:dyncomparison} (top),
where we have plotted the magnetic spectra measured in the bulk at
point 1 in VKS and in the simulation. As in the hydrodynamic case,
this long-term memory is associated with the evolution of the
large-scale modes. However, the magnetic field now appears to develop
a $1/f$ regime over a much wider range of scales (almost 2 decades)
than in the passively advected case. This indicates that the dynamo
field develops its own long-term dynamics, which increases the
correlation time scale as the magnetic field produced by the dynamo is
spatially correlated at the largest available scale in the box. As a 
result, the range of time scales for which long-term memory is observed is
substantially increased in the case of the self-sustained dynamo, which can also
be observed in the evolution of the position of the shear layer with time (see
Fig.~\ref{fig:dynshearlayer}). Interestingly, we also note that the
remarkable collapse of the experimental and numerical spectra in
Fig.~\ref{fig:dyncomparison} (top) is obtained when the frequency in
the simulation is normalized by $F_0=u_{rms}/2L$, with $L=\pi$ the
forcing scale of the TG simulation and not by $F_0=u_{rms}/L$ as was
the case for the passively advected magnetic field previously
discussed. This is related to the large scale correlation of the
dynamo field, which is now correlated over the entire simulation box
which is twice the forcing scale.

The existence of $1/f$ slow dynamics of magnetic fluctuations, related
to the large scale nature of magnetic field, seems to be a robust
behavior of MHD large-scale dynamos. It is consistently observed in
VKS experiment and TG simulations, which have similar flow geometries,
but different boundary conditions, MHD parameters (mostly in terms of
Prandtl numbers), and which have very likely different dynamo
generating mechanisms (VKS generates an axial dipole and is very
sensitive to electromagnetic boundary conditions, while TG generates
an equatorial dipole and operates in a homogeneous medium). A 
slow dynamics $1/f$ regime has also been reported in the Karslruhe
dynamo experiment, which has a different geometry, different
fluctuation levels, and different dynamo mechanism. In
Ref.~\cite{bib:Dmitruk2011} it is argued from the results of a large
number of numerical simulations that in systems that develop inverse
cascades (as, e.g., helical MHD turbulence) the development of $1/f$
noise and long-term memory is more robust, in the sense that it is
independent of details of the flow geometry and forcing. The present
results are in good agreement with these arguments, and confirms 
in experiments the previous numerical results.

An interesting observation can also be done regarding the dissipative
regime of the high frequency magnetic spectra. As observed in 
Fig.~\ref{fig:dyncomparison} (top), this regime is not resolved in the
experimental data, due to limitations in dynamical resolution of the
magnetic measurements when performed in the bulk of the flow. However,
the dissipation behavior can be better resolved from measurements in
the outer layer of sodium, in the vicinity of the copper shell (for
instance at point 3 in the VKS sketch in Fig.~\ref{fig:VKSetup}),
where no flow is present and magnetic dynamics results essentially
from diffusion processes of the turbulent magnetic fluctuations in the
bulk. Such measurements show that the far dissipation spectrum of
magnetic dynamo field decreases exponentially with the frequency 
$f$ (as an example, the bottom right panel of 
Fig.~\ref{fig:dyncomparison}) reproduces results from Fig.~16(c) 
in \cite{bib:monchaux2009_PoF}). Though we do not have an 
explanation for such an exponential spectral regime, we 
interestingly find that the same exponential behavior is 
captured by the simulation as shown in 
Fig.~\ref{fig:dyncomparison} (bottom left). TG simulations therefore 
are able to reproduce a large number of temporal spectral 
properties of turbulent magnetic fluctuations as measured in the 
VKS experiment.

\section{Conclusions}

We presented a comparison of time series of pointwise measurements of
the velocity and of the magnetic field, both from direct numerical 
simulations and from experiments of turbulence in conducting flows. 
Two configurations were considered: (1) A case with a weak externally
imposed magnetic field, in which induction generated by fluid 
motions can be used as a tracer of the velocity field (and mostly of the
large scale modes of the velocity field, as the magnetic Prandtl number 
considered in that case is small). And (2) a case with self-sustained
magnetic fields. The flow in the experiment corresponds to a von 
K\'arm\'an swirling flow between two counter-rotating impellers 
in a cylindrical vessel, while in the simulations the flow is
generated in a square box using periodic Taylor-Green forcing.

While experiments excel at providing time statistics of field
fluctuations with limited spatial information of the flow geometry,
numerical simulations tend to be performed at high resolution for
short times, thus providing substantial amounts of spatial information
with little time statistics. In the present study a different approach 
was used, considering low resolution simulations but extended for
very long times, so comparisons between experimental and numerical 
data can be done on the same grounds.

Good general agreement was found between spectral properties in 
simulations and experiments, both in the hydrodynamic (weak 
imposed magnetic field) and dynamo cases. Intermediate and 
high frequencies spectra from the experiments were found to be
well reproduced by the simulations. In particular, evidence of
long-term memory and $1/f$ noise at intermediate frequencies was found
in both experiments and simulations. In the hydrodynamic case, low
frequency magnetic spectra were found to deviate between the
experiment and the simulation. This can be attributed to different
induction processes operating in the bounded experiment and in 
the periodic simulations.

In the hydrodynamic regime, $1/f$ noise in the kinetic spectrum is 
a signature of the slow fluctuations of the largest structures of the 
flow, and particularly of the mid-plane shear-layer. As a consequence, 
this is only observed in the simulation when small wave numbers are
selected. This selection is naturally operated when considering 
magnetic fluctuations of a passively advected magnetic field in 
low Prandtl number situations, which are indeed found to exhibit 
$1/f$ behavior.

In the dynamo case, $1/f$ noise of magnetic field fluctuations is 
enhanced over a wider range of scales, that is attributable to the 
intrinsic large scale nature of the generated field. The remarkable 
collapse of simulations (at $\textrm{Pm}=0.5$) and experiments 
(at $\textrm{Pm}\approx 10^-5$) indicates that this may be a 
fundamental and robust property of large-scale MHD dynamos 
where slow fluctuations (both kinematic and magnetic) appear 
when the magnetic and velocity field become intimately coupled.
This last conclusion may be particular useful to extrapolate 
dynamical behaviors from simulations at $\textrm{Pm}\approx 1$ 
(a condition computationally favorable) to more realistic dynamos 
(natural or experimental) at small Prandtl regimes (which are 
computationally expensive to simulate). A particular example where 
this may be relevant is the case of dynamo reversals, which take 
place on very long time scales and which cannot be studied in 
simulations at low magnetic Prandtl number.

\begin{acknowledgments}
The authors acknowledge support from grant ECOS A08U02. 
PDM and PD acknowledge support from grants No.~PIP
11220090100825, UBACYT 20020110200359, and PICT 2011-1529 
and 2011-1626. The french authors acknowledge their colleagues of the VKS team with whom experimental data have been obtained. Their work was supported by ANR 08-0039-02.

\end{acknowledgments}

\bibliographystyle{apsrev}
\bibliography{./main}

\end{document}